# Performance Modeling of BitTorrent Peer-to-Peer File Sharing Networks

**Kunjie Xu**

**Graduate Telecommunications and Networking Program**
**School of Information Science**
**University of Pittsburgh**

**Abstract**

*BitTorrent is undoubtedly the most popular P2P file sharing application on today's Internet. The widespread popularity of BitTorrent has attracted a great deal of attention from networking researchers who conducted various performance studies on it. This paper presents a comprehensive survey of analytical performance modeling techniques for BitTorrent networks. The performance models examined in this study include deterministic models, Markov chain models, fluid flow models, and queuing network models. These models evaluate the performance metrics of BitTorrent networks at different regimes with various realistic factors considered. Furthermore, a comparative analysis is conducted on those modeling techniques in the aspects of complexity, accuracy, extensibility, and scalability.*

## 1 Introduction

Over the last decade, peer-to-peer paradigm has demonstrated itself to be an effective, scalable, and robust networking application to provide services for content sharing and personal communications. Instead of the traditional client-server model, peer-to-peer network combines the resource from all peers together and contribute to all peers in return, which is the essence of its success. Public attentions to peer-to-peer applications came first from highly popular file-sharing systems. BitTorrent (BT) [1] is today's one of the most popular peer-to-peer file sharing protocols used for distributing large amount of data. Bram Cohen designed the protocol in April 2001 and released the first implementation on July 2, 2001 [19]. BitTorrent technique was then quickly employed by Lindows, Blizzard and most Linux distributions. Recently, large scale web-service providers such as Facebook and Twitter used BitTorrent technique to push hundreds of megabytes new updates to all servers worldwide in an efficient way. The advantage of BitTorrent is delivering content in a scalable way, decreasing the load of central server, minimizing the distribution cost and reducing the download delay.

Generally, the P2P file sharing networks can be classified into four categories based on topology and degree of decentralization: those are (1) decentralized unstructured topology, (2) decentralized structured topology, (3) partially decentralized topology, and (4) centralized topology. In decentralized unstructured topology, such as Gnutella [2] and Freenet [3], all peers act as both server and client equally and the overlay network topology is freely formed by peers. For decentralized structured topology, like Chord [4] and CAN [5], each peer performs as both server and client simultaneously, but the overlay network topology is precisely controlled by a particular algorithm, such as distributed hash table (DHT) [6]. The partially decentralized topology, including FastTrack [7] and Brocade [8], has some super-nodes or super-peers that play a more important role than others. Centralized topology, such as Napster [9] and BitTorrent [1], is featured with a central server coordinating the interaction between peers. The centralized topology system is characterized by two attributes: centralized index and distributed download.

The central server only provides the directory service, and the file transfer is performed by distributed peers. There are two major advantages of centralized topology. Firstly, the resource management is easy. The central server maintains a central directory of the resources on the peers in the network. Every time peers join or leave the system, the resource index of that peer will be added or removed from central server. Secondly, the index and the resource discovery are efficient. When a peer requests the central server certain resources, the central server just look up its resource directory and then return the information about resource location immediately.

Besides the common features of peer-to-peer file sharing systems, the unique characteristics of BitTorrent make it quite popular to the Internet users and the networking research community. Generally, there are two regular steps required for any P2P file sharing systems: (1) locating the interested file resources among the participating peers in the network, and (2) downloading the resources from the peers being located. Following these two common steps, BitTorrent is characterized by its own features. First of all, BitTorrent does not provide a search function to locate the resource. Instead, a central server facilitates the progress of locating resource by looking up its centralized index list. Secondly, BitTorrent employs several innovative mechanisms, such as tit-for-tat (TFT) and rarest first (RF). These build-in mechanisms are extensively studied and affirmed to enhance the overall network performance and the individual download experience [10] [11] [12]. However, the great popularity of BitTorrent results in a significant increase in P2P file download traffic on the Internet. Numerous reports confirm the fact that BitTorrent has been the major P2P file sharing traffic in both Internet and LAN. For example, BitTorrent traffic is estimated to account for 18% of broadband Internet traffic [13] and 13-15% of the access link bandwidth at a residential university [14]. Moreover, BitTorrent applications present significant traffic-engineering challenges for Internet service providers (ISPs). The reason is that current implementation of BitTorrent ignores the underlying Internet topology or ISP link costs. As a result, a large amount of cross-ISP traffic is generated and the operating cost of ISP is increased accordingly [15]. In addition, another weak point of BitTorrent is the random neighbor selection strategy, which might lead to the inefficient usage of network resources. Hence, many researchers modified BitTorrent mechanisms and proposed BitTorrent-like protocols [16] [17] [18].

In this paper, we present a survey of performance modeling techniques focusing on BitTorrent P2P file sharing systems. In section II, BitTorrent system and its technical issues are introduced, for the purpose of placing further modeling discussion in proper context. Section III reviews various modeling techniques in details, including deterministic models, Markov chain models, fluid flow models, and queuing network models. A comparison of different models and the open issue are given in section IV. Finally, section V concludes this paper.

## 2 Overview of BitTorrent

### 2.1 What is BitTorrent

BitTorrent is a hybrid P2P file sharing system, where the interactions occur mostly among the peers and occasionally with a server for locating peers. Instead of downloading a file from a single source, BitTorrent protocol organizes peers into an overlay network named "*Torrent*" to download and upload the same file piece by piece among each other simultaneously. To distribute a file, a separate torrent has to be established in BitTorrent.

The peers in BitTorrent system is classified into two types: a *downloader* (*leecher*) and a *seed*. Downloaders are peers who only have a part (or none) of file, and seeds are peers who have a full file and stay online to serve others. Thus, seeds only perform uploading, while downloaders download pieces that they do not have and upload pieces they have. After a downloader finishes downloading a file, it may become a seed by staying online. To initiate a torrent, at least one initial seed is needed in the network to provide the entire content for download.

In addition to peers, tracker and web server also play significant roles in file distribution using BitTorrent. The *tracker* is a special infrastructure node which keeps track of all the peers which are (or recently were) active within a torrent (in the process of downloading the same file). Peers interact with the tracker using a simple protocol layered on top of HTTP. In the interaction, each peer sends information about the file it is downloading and the port number to the tracker, and then the tracker responds with a list of contact information of peers which are downloading the same file. The tracker does not participate in the actual distribution of the file, but only serves for the purpose of enabling peers to find each other. To distribute a file through a torrent, a static file with the extension .torrent is put on an ordinary web server. The .torrent contains information about the file name, the length, the hashing information of each piece, and the URL of a tracker.

To start downloading a file, Peer A has to obtain the corresponding .torrent file from a web server. Then, Peer A contacts the tracker and requests a list of IP/port pairs of other peers that are already to participate in the torrent. After the tracker receives a request, it will randomly select usually 50 peers from the set of all active ones within the torrent, and then sends the requesting Peer A a list of those 50 peers as potential neighbors. The random selection algorithm can generate a robust graph, which cannot be easily segmented after churn [19] [50] [47]. After receiving the list, Peer A will consider the peer from the list as its *neighbor*, if the peer can be connected to. Then Peer A will create a *peer set* consisting of all its neighbors, and try to download pieces from all peers in the set. Sometimes peers, however, do not have any piece they want or will not let them download. Strategies for peers preventing others downloading from themselves is called *chocking*, and it will be discussed later. When the number of its neighbors falls below 20 due to churn, Peer A will contact the tracker again for a new list.

In summary, there are five basic steps to establish a file sharing process in BitTorrent.
1) To download a file, Peer A first downloads the corresponding .torrent file from a web server.
2) Then, Peer A contacts the tracker for a list of active peers who is participating in the torrent.
3) Next, the tracker returns a list of peers who are involved in the torrent.
4) After that, Peer A adds all the connected peers from the list as its neighbors, and send the file piece request to each other
5) Finally, once the request is accepted, Peer A can exchange file piece with the neighbors.

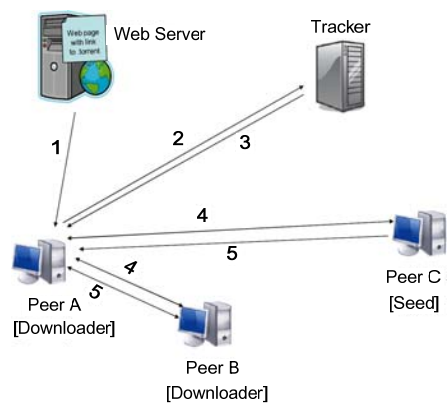

Fig.1. BitTorrent file sharing process [52].

## 2.2 Technical Issues in BitTorrent

*Pieces and Chunks*

In BitTorrent system, peers break a file into fixed size *pieces* (typically 256 KB each), and exchange pieces with each other. Once a peer downloads a completed piece, it computes the SHA1 hash of that piece and compares it with the value in the .torrent file. After the hash code is checked, the peer will report to all neighbors about its availability of that piece. In order to transfer data over TCP connection at full capacity, *pipelining* technique is used in BitTorrent. To achieve this, a piece is broken further into sub-pieces, typically of 16KB in size, which is named *chunks* or *blocks*. A number of request for sub-pieces, typically five, are kept pending in a pipeline. Each chunk arrival will trigger a request for another chunk. By this way, the file transfer connection is kept busy most of the time and the each piece is downloaded fast.

*Peer Sets*

In BitTorrent system, a peer can regularly upload to four peers plus one optimistic peer simultaneously. Based on the chocking algorithm, the peer selects the best peers to unchoke (upload), and chokes the rest of requesting peers. Chocking is a temporary refusal to upload, but the connection still exists. In other words, chocking stops uploading temporarily, but download can still occur. To better understand the chocking algorithm as the peer selection strategy, we have to distinguish five different peer sets in BitTorrent, which are (1) peer set, (2) interest set, (3) mutually interested set, (4) regular unchoke set, and (5) optimistic unchoke set. The *peer set* of Peer A is the set of all its neighbors. Only if Peer B has pieces which Peer A does not have, then Peer B is considered as Peer A's *interest set*. Similarly, if and only if Peer A also possess Peer B's interested pieces, then Peer B is regarded as Peer A's *mutually interest set*. Furthermore, if Peer B is unchoked by Peer A based on Peer A's basic chocking algorithm (tit-for-tat), then Peer B is a member of its *regular unchock set*. If Peer B is unchoked by Peer A by optimistic unchoking mechanism instead of TFT mechanism, then Peer B becomes a member of Peer A's *optimistic unchoke set*. The details of these mechanisms will be given next in this section. Fig.2 illustrates the relationship between different sets.

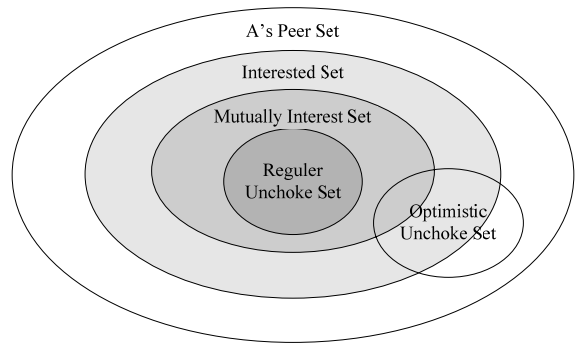

Fig.2. Peer sets in BitTorrent [52].

*Peer Selection Strategy*

The peer selection strategy is composed of four mechanisms: tit-for-tat (TFT), optimistic unchoking, anti-snubbing, and upload only. Different mechanisms are used in different scenarios, and they together form the basis of the chocking algorithm used by a peer. The motivation of these mechanisms is to improve the download experience of the peers who contribute to the system, and punish the free riders, who only download from other peers but never upload.

1) *Tit-for-tat*

Each peer uses the tit-for-tat mechanism to unchoke best four requesting peers who provide it highest download rate. Since the peer may receive requests from more than four peers, it has to choke the rest peers. By default, the tit-for-tat mechanism is run every 10 seconds by every peer, whereby the download rates are determined by the amount of data received over the recent 20 seconds. As shown in Fig. 3(a), Peer A unchokes best four peers B, D, F, and H to upload its pieces. The tit-for-tat mechanism enables the peers to provide an incentive to the neighboring peers who offer them high download rates, and penalizes the free riders.

2) *Optimistic Unchoking*

To discover new peers with better performance, a so-called optimistic unchoking is done additionally. Every 30 seconds, Peer A randomly unchokes a fifth peer who sends the request, and uploads to this peer. This peer has to be outside the peer set which Peer A is currently downloading from. For example, Peer A optimistically unchokes Peer J in Fig. 4. As a result, the remote Peer J might also unchoke Peer A in return during this 30 seconds. The benefit of optimistic unchoking is that it enables a peer to discover better neighboring peers to exchange pieces with, since unconnected neighbors may provide higher download rate than the currently unchoked peers. If a peer only uses the TFT mechanism, there will be no opportunity for detecting other peers who can provide higher uploading rate. Specially, this strategy is quite helpful for newly joined peers to get started. Hence, the tit-for-tat mechanism combined with the optimistic unchoking can periodically drop the peer with the lowest uploading rate in regular unchecked set, and keep the number of unchoked peers equal to five.

3) *Anti-snubbing*

Every 10 seconds, the tit-for-tat mechanism is executed and the unchoked peers are updated. If a peer is choked by the peer it was formerly downloading from, it regards itself being snubbed by that peer. Hence, the snubbed peer will choke that peer back as anti-snubbing. In another words, the peer will not upload to that peer anymore by regular unchoke. For example, after Peer A establishes a connection with Peer J by the optimistic unchoking, Peer A finds that Peer J provides higher download rate than Peer D. Hence, Peer A chokes Peer D according to the TFT mechanism. In return, Peer D also chokes Peer A based on the anti-snubbing. Meanwhile, Peer J becomes a member of Peer A's regular unchoke set. Then, Peer A randomly chooses another unconnected neighbor, say Peer K, in its optimistic unchoking period.

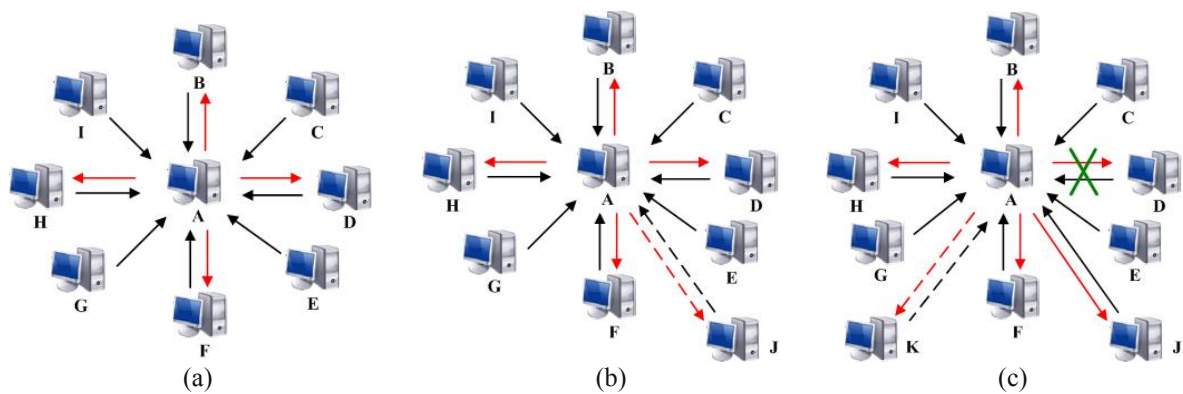

Fig.3. (a) Tit-for-tat mechanism; (b) Optimistic unchoking; (c) Anti-snubbing
(black-line: download; red-line: upload).

4) *Upload Only*

The TFT mechanism is not applicable for seeds, because seeds have nothing to download and they cannot select peers based on their uploading rates. Instead of TFT, seeds follow "upload only" mechanism to unchoke two types of peers. They first select the peers who utilize all their available upload capacity to serve others. Then, if the seeds are still capable of uploading to more peers, they will select the peers which no one else is uploading to at this time.

*Piece Selection Strategy*

In BitTorrent system, another successful strategy is the piece selection strategy. Basically, the piece selection strategy tells the peer which piece should be downloaded after current piece. A smart piece selection strategy can increase the availability of a full copy file in the network. In BitTorrent, the piece selection strategy is consisted of the following four mechanisms:

1) *Strict Priority (chunk-level selection)*

When some chunks (sub-pieces) are received from a specific piece, then the rest chunks of that piece are requested before chunks from any other pieces. In other words, peers will not request another piece until the current piece is downloaded completely. The reason of strict priority is that only a full copy of piece can be traded between peers.

2) *Rarest First (piece-level selection)*

Once peers finish downloading the current piece, it will select the next piece which is the fewest among its neighbors. This mechanism is named rarest first. In BitTorrent, each peer maintains a list about the number of each piece in its peer set to determine the rarest piece set, which includes the pieces that have the least number of copies in its neighbors. This rarest piece set is updated every time a copy of piece joins or removes from its neighbors. The benefit of this mechanism is to increase the chance of trading pieces between peers and efficiently distribute all the pieces in the network. Moreover, the rarest first mechanism enhances the file availability in the network, and reduces the risk of system death due to the random departure of seeds. From long run, the rarest first generates a uniform distribution of pieces among all online peers.

3) *Random First Piece (starting- point piece section)*

When a peer just joins a torrent, it has no piece at the very beginning. The rare pieces normally exist in one or few peers, and hence they will be downloaded to a peer slower than the regular pieces. The regular pieces are present on multiple peers, and different chunks could be downloaded from different peers at the same time. Therefore, instead of the rarest first mechanism, the first piece to download is selected randomly so that the whole piece could be obtained quickly. Then, this peer can trade with others, and the rarest first mechanism is initiated.

4) *Endgame Mode (end-point chunk selection)*

At the end of download, a requested piece is sometimes transferred with a very slow rate. This is not a problem in the middle of download, but could be a matter to delay a peer's finish. To solve this problem, a peer sends request to all of its neighbors for all chunks of the last piece. As soon as a chunk is received, the request of that chunk is cancelled to prevent redundant downloads. In this way, the last piece of a file is downloaded quickly.

In summary, both the peer selection strategy and the piece selection strategy play significant roles on the performance of BitTorrent system. In particular, peer selection strategy enhances the service capacity of the system and reduces the download delay of each peer. The piece selection strategy guarantees that each peer can find their interested pieces from its neighbors quickly, and improves the file availability and the lifetime of the network.

# 3 Performance Modeling

## 3.1 Performance Metrics

The peer-to-peer system is a distributed network that partitions tasks or workloads between peers, and provides ad hoc collaborations among peers. Due to heterogeneous bandwidth, unexpected user behavior, free rider, and churn, it is not straightforward to examine the performance of such a system. In the past few years, considerable effort has been spent in research on the performance modeling of P2P file sharing systems. In the following, we list the major performance metrics, and associate a short discussion with each of them.

*Service Capacity (C)*

From system perspective, service capacity is the overall achievable throughput the system can offer the downloaders in a torrent. As a system-level measure of resources, the service capacity takes into account the overall effective upload bandwidth from both seeds and downloaders. "Effective" here reflects the fact that seeds can upload at their available bandwidth, while downloaders may only be able to upload at a fraction of their available bandwidth because they do not have the completed copy of a file.

*Download Throughput per Peer ($\alpha$)*

From users' perspective, download throughput per peer is considered as the service capacity to each peer. It might be roughly estimated as the aggregated upload service capacity normalized by the number of downloaders.

*File Download Latency ($d$)*

File download latency is the time from the file request being sent out until the file being downloaded completely. It is consist of two components: the query search time and the file transfer time. In the P2P paradigm, any file distribution application implemented requires two different functions to be supported: (1) a search function that enables peers to locate the content that they are interested in among the participating peers, and (2) a download function that enables peers to download the content once it is located. For BitTorrent system, a central server contains an index of all the files that the nodes share in the P2P system. In such a centralized architecture, the search time for a file query is mainly the lookup time to retrieve the information in the central server, which is much smaller than the time required to download all pieces of the file from peers. Thus, the expected query search time is neglected in most of the modeling works on the centralized architecture.

*File Availability / Torrent Lifetime ($T$)*

A file is available only when a peer can download all the pieces needed from seeds or downloaders in the system. If there is always at least one seed in the system, then the file is guaranteed to be available all the time. But in reality, the seeds may want to minimize the time of staying in the system and the peers may choose to depart from the system once they obtain the whole file, or they may abort in the middle of the file download due to unexpected user behavior or network failure. Thus, the system may lose some pieces due to the departure of downloaders and seeds, and the remaining download processes will never finish. In addition, the torrent lifetime of a specific file can be directly related to the availability of that file, since torrent lifetime is the period in which the torrent could provide a complete file from all the online peers. If the pieces kept by the peers are incomplete, then the torrent is dead.

*Network Stability*

In a stable P2P file sharing network, all download requests will be served and cleared in finite time, if the average workload does not exceed the average system service capacity. Only if the number of peers and the performance of each peer are relatively stable, then we say the system enters "steady state". Hence, the stability of a P2P network is a fundamental problem for any serious steady state performance analysis. The stability issue in P2P file sharing networks has recently been investigated by using queuing theory [20] and control theory [21] [22].

## 3.2 Key Factors to Effect Performance

*Churn*

The collective effect created by the independent arrival and departure of thousands or millions of peers is called churn, which is an inherent property of P2P systems. A peer can randomly join or leave the system when the user wants to start or exit the application. Given the cooperative nature of P2P networking, intermittent connectivity between peers may lead to severe performance degradation in terms of file downloading. Hence, the user driven dynamics of peer participation must be taken into account in the performance analysis of P2P system.

*Free Rider*

A free rider is a user who obtains resource from the network without contributing to it, either selfishly or unintentionally. This behavior leads to a tangible level of unfairness to the remaining participants who contribute to the system. If a high enough proportion of users are free riders, the download performance will degrade for those who do contribute. A substantial drop in performance may drive some contributors to leave the system altogether, which worsens the situation further for the remaining peers and leads to a vicious circle. Hence, free riding has been considered as the major threat to the performance of P2P networks.

*Bandwidth Issue*

The heterogeneity of peer bandwidth seriously impacts the performance of the whole network and the individual peer. The diversity of peer bandwidth comes from two reasons. First of all, there is a significant heterogeneity of link access capacity in today's infrastructure. Measurement results from [36] reveals 8% of users connected by means of modem (64 Kbps or less), 60% using broadband connections (DSL, cable, T1, T3), 30% having very high bandwidth (at least 3 Mbps) connections. Obviously, low bandwidth access can directly degrade the peer's downloading experience. Secondly, numerous bandwidth allocation strategies have been proposed to improve the performance of BitTorrent network or control the BitTorrent traffic on the Internet. For example, to improve the incentive mechanism of BitTorrent, different bandwidth capacities should be allocated to peers based on their various contributions. Peers who provide more service to others will be granted a higher download priority and bandwidth than peers who provide less service. In addition, the extensive use of P2P file exchange causes congestion and performance deterioration of today's Internet. The traffic control policies coping with excessive bandwidth consumption by P2P traffic are implemented on the ISP gateway, and they significantly affect the download performance of subnet users.

Besides the above factors, the build-in features of BitTorrent also improve the file downloading performance, such as the multi-piece download, the piece selection strategy and the piece selection strategy. All of these factors are interrelated in complex ways. For example, a smart peer or piece selection scheme may favor peers that are likely to subsequently stay as servers for the file and thus affect the statistics of churn. An incentive bandwidth allocation strategy to prevent free rider behavior can encourage seeds to stay in the network longer, and may influence the churn also.

### 3.3 Performance Modeling Techniques

Several analytical models of BitTorrent P2P file sharing system have been proposed in the literature. In this paper, we classify the majority of modeling techniques in four catalogs: (1) deterministic models, (2) Markov chain models, (3) fluid flow models, and (4) queuing network models. Analytical models are supposed to clearly reflect the effects of different parameters on BitTorrent network performance. They permit efficient and detailed exploration of the parameter space to evaluate the effect of not just a single parameter variation, but also the combined effect of several parameters variations. However, many models are based on the assumptions that peers are aware of the global information about the download state or the upload/download bandwidth of all peers. Also, many models simplify assumptions on the underlying network topology, and on the arrivals and departures of peers. In general, the typical parameters that characterize the system evolution include:

- $x$, the number of downloaders (also known as leeches) in the system.
- $y$, the number of seeds in the system.
- $\lambda$, the arrival rate of new requests.
- $\mu$, the upload bandwidth of a peer.
- $c$, the download bandwidth of a peer.
- $\theta$, the rate at which peers abort their download.
- $\gamma$, the rate at which seeds leave the system.
- $\eta$, the effectiveness of file sharing, ranged within [0, 1].

Based on the above typical parameter settings, there are several extensions in different analytical models. For example, the upload and download bandwidth of peers could be homogeneous, where all peers have the same upload and download rate; or heterogeneous, where peers have different uploading and download rates. Another example is that people usually consider the sharing of a single piece as an independent process of other pieces and estimated the life time of the file sharing system by the disappearance of a single piece. But in reality, the dynamics of a single piece depends on the other pieces as well based on BitTorrent piece selection strategy. Hence, the performance of a more complex system with multiple-piece download should be investigated. One more extension in some models is to keep track of the different stages of download that a peer may exist at any point of time. This typically involves dividing the downloading peers $x$ into different subsets $x_1, x_2, x_3, x_4, x_5$, and etc. representing the different completion stages of download. Similarly, some models separate the peers $x$ into $x_n$ and $x_f$ corresponding to non-free riders and free riders, respectively.

From the real trace measurement, it has been observed that a BitTorrent system has two phases, a transient phase and a stationary phase, from the viewpoint of download throughput per peer for a given file [23]. This real trace measurement result is shown in Fig. 4. The "transient phase" represents the initial system performance in response to a large burst of requests for a popular file when it is first introduced, or due to periodic variations in load. During this period, only a relatively small number of peers have pieces of file and exchange with each other. As replication proceeds, however, more and more peers have the whole or parts of the file and serve the burst of requests, which enable an exponential growth in the service capacity of the system. Once the request rate becomes stable, the network enters a "stationary phase" where the performance of the whole network and each peer becomes stable. In the following parts of this section, various performance modeling techniques with different applied regime will be described in details.

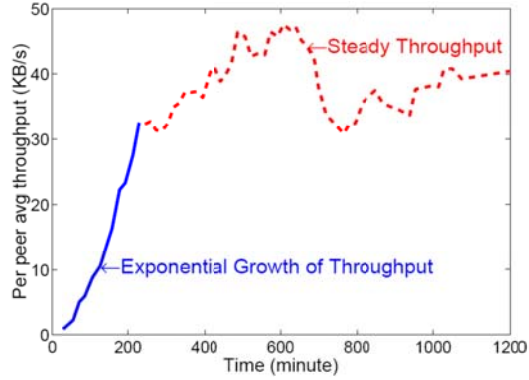

Fig.4. Two-phases in the evolution of the download throughput per peer versus time after a single file is introduced into a BitTorrent network [23].

### 3.3.1 Deterministic Model

The deterministic model [23] [24] [25] is typically applied to investigate the performance in the transient regime. Such a transient regime occurs when a large number, say $n$, burst of roughly concurrent requests arrive in the P2P network with only a limited number of peers, say one, capable of serving them. Such a scenario may come up due to the first introduction of a popular file or a periodic request pattern associated with a file. This file replication procedure is illustrated in Fig. 5, can be best explained by the deterministic model.

We suppose that there are $n = 2^k$ peers requesting a file in the network, with homogeneous upload bandwidth capacity $\mu = b$ and unconstrained download capacity $c = \infty$. The whole file is assumed as one piece with size $s$ bits, and a peer can only serve others once it has fully downloaded the file (single piece download). In order to serve $n$ requests, $ns$ bits have to be replicated and transferred to all peers. The best strategy is to first serve one user with rate $b$, so that the system service capacity grows to $2b$, and then these two peers serve other peers, until all the $n$ users get served, as illustrated in Fig. 5. Based on this strategy, peers will complete downloading the file every $\tau = s/b$ seconds, and then serve others. Hence, after every $\tau$ seconds, the system service capacity is doubled, leading to an exponential growth of $2^{t/\tau}$ in the unit of number of peers available to serve others. Following this dynamics, $n$ peers will be served by time $\log_2 n = k$. Therefore, we compute the average download latency $\bar{d}$ experienced by a peer as

$$\bar{d} = \frac{1}{n}\sum_{j=1}^{n} d_j = \sum_{i=0}^{k-1} 2^{i-k}\tau(i+1) = k\tau - \frac{n-1}{n}\tau = \tau\left(\log_2 n - \frac{n-1}{n}\right) \approx \tau \log_2 n. \quad (1)$$

Let $d_j$ denote the delay experienced by the $j$th peer to complete downloading, and note that $2^{i-k}n$ peers complete service at time $i+1$. Hence, for a P2P network with an initial burst of n requests, the average download delay per peer scales as $\log_2 n$ which is favorable relative to the linear scaling of $n$ one would obtain for a system with a fixed set of servers.

To further reduce the download delay, multi-piece download is implemented in many P2P file sharing systems, such as BitTorrent. Suppose the file is divided into $m$ pieces with identical size, and peers serve those pieces as soon as they are downloaded. This essentially enables the file transmissions pipelining, and fasten the download delay by a factor of $1/m$ [23] [25].

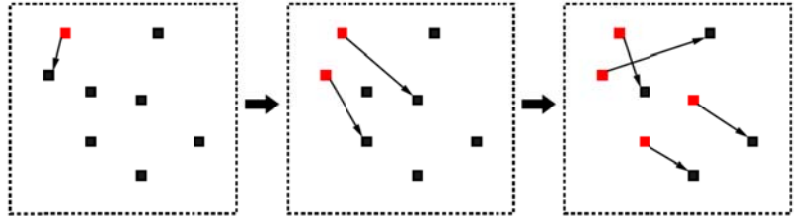

Fig.5. File sharing in P2P system [23].

The above idealized deterministic model illustrates the basic dynamics of file download latency one might expect from a BitTorrent system during the transient regime. More rigorous formulations for this type of problem are also discussed in [26] [27] [53]. And they provide a closed form solution with similar logarithm scaling property of the file dissemination delay. In summary, the deterministic model provides us a basic idea about the scalability properties of BitTorrent systems during the transient regime, in the sense that average download latency grows logarithmically as a function of the burst size (the total number of requesting peers).

### 3.3.2 Markov Chain Model

**A) *Typical Model***

The Markov chain model [23] is used for the steady state analysis of BitTorrent P2P file sharing system when the demand becomes fairly stationary and sustained after an initial transient phase. At the stable state, the arrival of the peers is not in a burst, but could be viewed as a Poisson process with rate $\lambda$. The system state is a pair $(x, y)$, where $x$ represents the number of downloaders, and $y$ denotes the number of seeds still serving the system. They further assume the file is partitioned into pieces and multi-piece download is allowed, so the peers in the process of download can provide their available pieces to others. Hence both seeds and downloaders contribute to the system service capacity. The contribution of the downloaders is only a fraction of a seed which has the full file. After the peers download the completed file, they are eligible to be the seeds to fully serve others. Hence, the total service capacity is denoted by $\mu(\eta x + y)$. However, the seeds may leave the system at rate $\gamma$. The evolution for the states of this system can be described by a continuous time Markov chain model with matrix $Q$

$$
\begin{aligned}
q((x,y),(x+1,y)) &= \lambda & &\text{new arrival} \\
q((x,y),(x-1,y+1)) &= \mu(\eta x + y) & &\text{serve a peer} \\
q((x,y),(x,y-1)) &= \gamma y & &\text{exit systems.}
\end{aligned} \quad (2)
$$

Furthermore, the service capacity of the P2P system and the file download delay per peer would be the sum of uploading throughput of both downloaders and seeds, which is

$$C = \mu\eta x + \mu y. \quad (3)$$

Then, we derive the download throughput per peer in steady state as

$$\alpha = \frac{C}{x} = \frac{\mu\eta x + \mu y}{x}. \quad (4)$$

In centralized architecture, the expected query search time could be neglected. Hence, the steady state file download latency can be directly related to the download throughput per peer as

$$d = \frac{1}{\alpha} = \frac{x}{\mu\eta x + \mu y}. \tag{5}$$

The states $(x, y)$ with $y = 0$ in the Markov chain are absorbing states. When all the seeds leave the system, the complete set of file pieces in the network cannot be guaranteed and the death of the whole file sharing process is irrevocable. Therefore, the absorption time starting from the initial state $(0, 1)$ can be viewed as the life time of the system [28] [29]. Thus, the mean time to absorption, i.e. the life time of the system, is the sum of the time spent in the non-absorbing states:

$$T = \sum_{(x,y):y>0} z_{(x,y)} = \sum_{(x,y):y>0} \int_0^\infty \pi_{(x,y)}(t) \, dt. \tag{6}$$

Where $z(x, y)$ denotes the mean time spent in state $(x, y)$ from the beginning ($t = 0$) to the absorption of the system, and $\pi_{(x,y)}(t)$ is the probability of state $(x, y)$ at time $t$.

**B)** *Model Extensions*

*Download and upload bandwidth constraints*

When the download/upload bandwidth constraints in the real systems are considered [21] [28] [29], the total service rate of the system becomes $\min\{cx, \mu(\eta x + y)\}$. In other words, the transition rate $q((x,y),(x-1,y+1))$ in Markov chain transition matrix $Q$ turns out to be $\min\{cx, \mu(\eta x + y)\}$. If $cx < \mu(\eta x + y)$, the downloaders cannot use all upload capacity and the download bandwidth is the constraint. Otherwise, the upload bandwidth becomes the constraint.

*Multiclass peers*

In [30], the Markov model is extended to study the distribution of peers in different states of download completion. Towards this end, they explicitly represent the peers as one of several states $S_0, S_1, \ldots, S_{N-1}$ with $[\frac{0}{N}, \frac{1}{N}), [\frac{1}{N}, \frac{2}{N}], \ldots, [\frac{N-1}{N}, 1)$ portions of the file respectively, and $S_N$ is the state for seeds. So the peer state behavior can be viewed as a continuous time Markov chain. Then they derive the distribution of peers in each state of download completion. The distribution is observed to be a U-shaped curve in which the peers are more concentrated at both sides which are the states with very small or large portions of the file. From the model, they find that seeds departure rate and downloaders abort rate can seriously affect this distribution.

*Multiple pieces download*

An extended Markov chain model with multiple pieces download is proposed in [31]. In particular, they consider a system splitting the file from one piece into two pieces. They conclude that dividing the file into pieces improves the performance of the P2P file sharing systems in terms of file availability and average download time. The biggest improvement is done when the file is divided from one piece into two pieces. Whether the number of the pieces is 10 or 100 does not have such a significant influence on the performance anymore, since increasing the number of the pieces also raises the overhead costs and delays.

*Impact of free rider*

When the impact of free rider is considered [32], the network performance will be degraded due to the selfish behavior of free riders. The ratio of free rider population and overall population is assumed to be $f$ in the network. Since free rider just downloads from others but never contributes to the network, the file sharing effectiveness of free riders is equal to 0. Hence the average effectiveness of file sharing per peer is reduced from $\eta$ to $\eta(1-f)$. In addition, after free rider leaves the system immediately after completing his download, the average rate at which seeds leave the system is increased from $\gamma$ to $\gamma/(1-f)$.

*Generalized Markov chain model*

A generic Markov chain model is introduced [20] to model the P2P file sharing systems with different policies. They assume that both job and server arrive and depart randomly, and the server dynamics may or may not correlate to the job dynamics. The "job" here is equivalent to the "downloaders", since the download request from the peers is regarded as jobs here. However, the "server" here is not just "seed", since we know "downloader" also contribute to the network by serving others and regard it as "server" as well. To represent queuing models for P2P service systems in which both job and server dynamically arrive and depart, they use two dimensional Markov chain to model the P2P network with notation A/B/(C/E)/POLICY, where A/B and C/E represent the dynamics of job and server, respectively. Moreover, the service policy (POLICY) includes (1) *FCFS* (First-Come-First-Served), which represents jobs are served in the arrival order by all the current servers in the system; (2) *PS(k)* (*k*-Processor-Sharing), which indicates each jobs in the system is served by no more than *k* servers simultaneously. In real systems, this *k*-processor-sharing constraint is from the downlink capacity limitation of peers. To represent systems with different job and server dynamics, the notations used for arrival distribution (A or C) and lifetime distribution (B or E) could follow (1) Poisson process *M*, (2) deterministic process *D*, and (3) general or arbitrary process *G*. In addition, we notice that the server dynamics might be correlated with the job dynamics, since every downloader in the network also acts as a server. There are two possible correlation cases. The first case is when the server arrival and lifetime distributions are identical to the job arrival and lifetime distributions, then the notation "-" is used for C and E. The second case is when server lifetime equals to the summation of job lifetime and an extra period of time following exponential or arbitrary distribution, the notations +*M*/+*G* are used to describe the server lifetime distribution E. In this case, after the peer finishes downloading a complete file, it stays online as a seed for a while to serve others. Hence, four classes of P2P systems are analyzed, those are (1) *M/M/(M/M)*/FCFS, (2) *M/M/(M/M)*/PS(k), (3) *M/M/(-/-)*, (4) *M/M/(-/+G)*. In this work, the stability conditions are given for those four classes of systems based on the above Markov chain models. They show that if the average workload does not exceed the average system service capacity, then the P2P system is stable, i.e. all arriving job will be served and cleared in finite time.

### 3.3.3 Fluid Flow Model

A more general approach to evaluate the performance of BitTorrent networks is to use the fluid flow modeling technique, which can model the network behavior during any nonstationary periods. Fluid flow models are quite general in nature and can also model the non-Markov queues in different types of networks [48] [49] [50]. Due to churn, unexpected behavior of each peer, and bursty /nonstationary traffic in BitTorrent networks, high variability likely dominates the network performance. Thus, one would expect that fluid flow models can appropriately describe the dynamic performance of BitTorrent network.

**A) *Typical Model***

A simple fluid flow model is first derived in [21] to represent a dynamic systems incorporated with realistic factors such as downloaders departure (notated by *θ*), seeds departure (notated by γ), effectiveness of file sharing (notated by η), and download bandwidth constraint of a peer (notated by *c*). This model is associated with one file in size one (single piece file). The download bandwidth constraint is considered, and the time-varying service capacity *C*(*t*) (total uploading rate) is the minimum value of downloading and uploading bandwidth capacity in the network, i.e. $C(t) = \min\{cx(t), \mu(\eta x(t) + y(t))\}$. The continuous-time quantities (fluid) in the system are the number of downloaders *x*(*t*), and the number of seeds *y*(*t*). This fluid flow model for the evolution of the peers number (downloaders and seeds) is give by

$$\frac{dx(t)}{dt} = \lambda - \theta x(t) - \min\{cx(t), \mu(\eta x(t) + y(t))\}, \quad (7)$$

$$\frac{dy(t)}{dt} = \min\{cx(t), \mu(\eta x(t) + y(t))\} - \gamma y(t), \quad (8)$$

In addition, we have the time-varying download throughput per peer *α*(*t*) as

$$a(t) = \frac{C(t)}{x(t)} = \frac{\min\{cx(t), \mu(\eta x(t) + y(t))\}}{x(t)}, \quad (9)$$

This fluid flow model is analyzed in steady state in [21], which is conditioned by

$$\frac{dx(t)}{dt} = \frac{dy(t)}{dt} = 0, \quad (10)$$

Then, the average number of downloaders and seeds in steady state are given

$$\bar{x} = \frac{\lambda}{\beta(1+\frac{\theta}{\beta})}, \quad \bar{y} = \frac{\lambda}{\gamma(1+\frac{\theta}{\beta})}. \quad (11)$$

With $\frac{1}{\beta} = \max\{\frac{1}{c}, \frac{1}{\eta}(\frac{1}{\mu} - \frac{1}{\gamma})\}$. Further, the average file download latency $\bar{d}$ is given by:

$$\bar{d} = \frac{1}{\theta + \beta}. \quad (12)$$

It is observe that the average download time $\bar{d}$ is not related to the request arrival rate *λ*, which means BitTorrent system scales very well. Furthermore, if the seed leaving rate *γ* is smaller than *μ*, the download bandwidth *c* determines the network performance. The author of [21] further investigates the stability issue in [22]. They consider BitTorrent network as a switched linear system and prove that such a system is always globally stable by using Lyapunov function [33].

**B)** *Model Extensions*

*Exponential Decreasing Peer Arrival Rate*

Based on the extensive measurements and trace analysis in [34] [35], they argue that the peer arrival process does not simply follow the Poisson distribution. Instead, they conclude that the peer arrival rate decreases exponentially, denoted as $\lambda(t) = \lambda_0 e^{-t/\tau}$. Here, $\lambda_0$ is the initial arrival rate when the torrent starts, and *τ* is the attenuation parameter of the file popularity. Then, they extend the fluid flow model in [21] by replacing λ with the exponential decreasing peer arrival rate λ(*t*). From the new model, it is found that both peers and seeds increase exponentially first, but then decrease exponentially at a slower rate. Such results fit the trace very well, especially for the torrents with large population. From the model, they finally conclude that due to the exponentially decreasing peer arrival rate and the limited online time of seeds in a torrent, the service availability of the corresponding file is degraded quickly, and eventually it is hard to locate and download this file. In other words, the seeds departures disenable the torrent to keep up with the peer service demand, and the torrent dies eventually. In addition, they conclude that the random arrival and departure of downloaders and seeds cause the performance to fluctuate significantly over the torrent lifetime, especially for small-size torrents.

*Heterogeneous Link Capacity*

The effect of heterogeneous link capacities on the performance of BitTorrent file sharing systems is addressed in [36]. To keep the heterogeneity simple, they consider the peers with only two possible link capacities: high speed peers and low speed peers. In addition, the links are assumed to be symmetric, i.e., the upstream rate is the same as the downstream rate. Based on the extended model, they analyze the effects of bandwidth heterogeneity on file transfer dynamics in details. Specially, they compare the performance of heterogeneous networks and the equivalent homogeneous networks under difference scenarios. Their numerical results show that the bandwidth heterogeneity can have a positive effect on content propagation among peers if the appropriate network scenarios are chosen.

*Peer Heterogeneous Behavior*

The authors in [37] focus on the behavior of the different peers during download procedure. The behavior of downloaders is divided into two types: high tolerance downloaders and low tolerance downloaders. For the former one, after peers enter the system, they usually stay throughout the entire download procedure. In other words, they have a high tolerance for download latency. The authors explicitly define that the departure rate of high tolerance downloaders ($\theta_a$) is lower than their download rate at the maximum capacity, i.e., $\theta_a < c$. On the contrary, the low tolerance downloaders are the peers that have very little tolerance to download latency. The explicit definition is that the departure rate of low tolerance downloaders ($\theta_b$) is higher than their download rate at the maximum capacity, i.e., $\theta_b > c$. Based on this two-population model, a priority-based peer selection scheme is proposed for the BitTorrent network when the resources in the system are scarce. This scheme is to first serve the high tolerance downloaders before the low tolerance downloaders. In other words, the peers who stay in the system longer are rewarded by being served first. The results show that the priority scheme is well suited and improve the download performance of the system whenever the environment is adverse in terms of insufficient upload bandwidth.

*Bandwidth Allocation Strategy*

In practice, peers could have diverse bandwidth, such as Ethernet access, dial-up modem access and broadband access. In [38], a general multiclass model is presented for heterogeneous peers with different access bandwidth. In particular they carefully consider a system with two classes of peers, distinguished by their different upload and download speeds. Furthermore, it is reasonable for BitTorrent system to provide differential service by allocating more upload bandwidth to the first-class peers, who pay more than the second-class peers. Thus, a single parameter $α_i$ is defined as the fraction of peer's upload bandwidth allocated to the class $i$ peers and this parameter is involved in their fluid flow model. Then they apply this model to account for two specific problems: service differentiation and bandwidth diversity. For the service differentiation problem, they show how to choose the value of $α_i$ to achieve a target Quality of Service (QoS) ratio (file download delay ratio between two classes of peers). For the bandwidth diversity problem, they illustrate how to determine the value of *α*ᵢ to minimize the maximum average download time of both classes.

*Multiclass Peers*

In [39], the authors divide the peers based on two different criteria. For the first dividing criterion, they separate the peers into three types: downloaders that have a few pieces of the file, downloaders that have most pieces, and seeds. They define the connection probability $\rho(\leq 1)$ as the probability that a peer maintains connectivity with another peer in the torrent. The file download time and the file availability are analyzed in BitTorrent systems with this model. They also obtained the closed-from solutions for the average number of seeds and downloaders, the average download time and the steady state service capacity of the system. From the closed-form solutions, the effect of various system parameters including peer arrival rate, seed departure rate, peer connection probability and transmission bandwidth on the performance measures is illustrated explicitly. Moreover, the sensitivity analysis of various performance metrics is carried out. This diving criterion is also applied in [40]. For the second dividing criterion, they assume two classes of peers: peers with publicly routable IP address, and peers behind firewall. The impact of firewall on the system is discussed, and the results clearly show that those peers not behind firewalls play an important role in determining the overall system performance.

*Impact of Free Rider*

In order to capture the impact of free-riding, an fluid flow model with two classes of peers is introduced in [41], those are free riders $x_f(t)$ and non-free riders $x_n(t)$. To simplify the model, both free riders and non-free riders are assumed to depart the system immediately after they have finished their download and have all piece of the sharing file, i.e. $\gamma_f \to \infty$, $\gamma_n \to \infty$. Hence, there is no seed existing in the system. Based on their model, the results show that tit-for-tat peer selection strategy is successful to guard free-riding influence to the performance of BitTorrent system without seeds. However, peers might stay in the system for a while after they finish the download in practice, and hence seeds have to be taken into account. As a result, in their following work [42] [43], they examine the free-riding impact on BitTorrent system with seeds and show that the free rider can take a great deal of advantage from seeds. The reason is that BitTorrent does not employ any effective mechanism for seeds to effectively guard against free-riding. Therefore, a seed bandwidth allocation strategy is proposed to reduce the impact of free-riding by considering the diverse contributions from downloaders.

### 3.3.4   Queuing Network Model

Queueing networks are suitable for representing the structure of various systems with a large number of resources. Queueing network modeling is a particular approach to model the computer system consisting of numerous service stations connected with each other. A station, i.e., a node, in the network represents a resource in the real system. Jobs in principle can be transferred between any two nodes of the network. In BitTorrent P2P file sharing systems, queuing network models could be used to model the structure of many peers connected as a network, or the structure of a sequence of behaviors of peers at different operating stages by abstraction. Different queuing network models emphasize on the various performance metrics, as illustrated below.

*File Transfer Latency Modeling*

In [51], an open queuing model is developed to evaluate the P2P file transfer latency in the network consisting of end peers and core routers in both centralized and decentralized P2P systems. The total file transfer delay is given by the sum of the (1) query search time, (2) peer level delay, i.e., the transmission time of the file being downloaded, and (3) core network delay, i.e., queuing delay at the intermediate routers. In centralized architectures, like BitTorrent system, a central server contains an index of all the files that the nodes in the P2P system share, and the search time for a query is the average lookup time of that central server to retrieve the information. Thus, the expected query search time is quite small compared with the file download time, and can be treated as a constant value. They model each peer as an $M/G/1/K$ processor sharing (PS) queue to evaluate the peer processing delay, while model each router as a $GI/G/1$ queue to estimate the core network delay. This model accounts for a number of factors in real P2P systems, such as heterogeneous file popularity/sizes, number of simultaneous downloads in peer settings, different link rates, diverse physical topologies, and churn.

*Peers Behavior Modeling*
In [45], a closed queueing network is proposed to model the peer's behavior for the general P2P file sharing networks. The peer's behavior consists of a series of phases connected with each other, which are user connection, content searching and query processing, file transfer, and user disconnection. This queuing network model is quite general, and can capture three different architectures (centralized topology, decentralized unstructured topology, and decentralized structured topology) and two class of peers (free riders and non-free riders). Despite their architectural difference, three important functions are performed in common by all peers: (1) maintenance of the infrastructure of the peer-to-peer system, (2) handling queries, and (3) file transfers. The manners in which the first two functions are accomplished significantly differ among three architectures. In the centralized topology, the central server handles all infrastructure maintenance and query processing functions. In contrast, these two functions are distributed among all the peers in the two decentralized architectures. However, the third function about file transfer is performed similarly in all three architectures. Thus, this modeling work can be applied to the performance analysis of BitTorrent system. Since this paper focuses on peer's behavior modeling, the related factors such as user behavior in terms of connection and disconnection policies, content and query popularity distribution, and free riders are taken in account. The performance metrics investigated in this work include service capacity of the system, average download throughput per peer, and average file download latency per peer.

## 4 Comparison and Open Issues

### 4.1 Comparison

Table I gives a summary of the various performance modeling techniques of BitTorrent P2P file sharing systems discussed in this paper. In this table, the typical models and the possible extensions by considering more realistic factors are shown in the first two columns for deterministic models, Markov chain models, and fluid flow models. However, since the works of queuing network models are dispersive, each work has its own queuing network formulation and there are no typical or extended models. The third column gives the performance metrics investigated in different models, and those metrics are defined in section 3.1.

In Table II, these four modeling techniques are compared in the aspects of applied regime, complexity, accuracy, extensibility, and scalability associated with the solution. We will carefully discuss these models in each of those aspects as follows.

Deterministic models are applied to the initial transient regime once a file is introduced into the network by a seed. At this time, a large burst of nearly concurrent requests enter the system but only few peers can serve. In this case, the exponential growth of system service capacity can be described as a deterministic replication process, which is easy to derive. However, the deterministic model is somewhat idealized and far away from reality. There is no consideration about various link capacity, heterogeneous peer behavior, random peer arrival process and free riding. In addition, the possible parameters that could be added on this model for extension are quite limited. In the deterministic model, since the number of peers $n$ is considered as a variable to determine the file download latency, the scalability of this model is not an issue.

Once the arrival rate of file request becomes stable, the system goes into steady state where system service capacity and throughput per peer is stationary. Hence, Markov chains are appropriate modeling tools to evaluate the steady state performance. Based on the trace measurement results in [23], the authors conclude that only if the file is popular enough to attract a large amount of peers, then the system will reach the steady state and hence Markov chain models are applicable. For the file with less popularity, the network performance is quite unpredictable. This big variance of the performance results from the fact that the number of peers is small and they join or leave the system randomly. To improve its accuracy, the states in Markov chain could be extended, and the transition rate could involve more parameters. When more peers join the network, more states are required in the Markov chain, which brings about the issue of its limited scalability. For the steady state Markov chains with special parameter matrix structures, the closed-form solution methods are sufficient. However, for the steady state Markov chains with a general parameter matrix structure, we need to resort to numerical methods. Generally, there are two classes of numerical methods to solve them: direct methods and iterative methods. Direct methods are subject to the accumulation of round-off errors and applicable to medium-sized (around 500 states) models, while convergence is not always guaranteed in iteration methods [54].

Fluid flow models are proposed to describe the time-varying system performance, since churn and dynamic burst traffic could make the system nonstationary quite often. Fluid flow models are quite general in nature and can also modeled non-Markovian queues. The accuracy of the fluid flow model is evaluated by both simulation [21] and measurement [35]. From their results, we see that the fluid flow model is a good approximation of the system due to its time-varying nature. In addition, fluid flow models are extendable by including more time-varying parameters to capture more details of network dynamics. Fluid flow models are applicable to large-scale BitTorrent P2P networks. Regardless of the number of peers, fluid flow models focus on the system-level analysis, and the typical model is always consisted of two differential equations, as shown in (7) and (8). Each of those two differential equations describes the change rate of the number of downloaders and seeds, respectively. Even though this model can be extended into more differential equations by considering multi-class peers, the computation complexity of solving them with the Runge-Kutta algorithm increases linearly with the number of differential equations [44].

Queuing network models are also applicable to the performance analysis of steady state system, and have been popular mathematical tools to model many real-world computer networks. For P2P network, the station/node in the queuing network could be a peer or a phase of peer behavior by abstraction. Hence, queuing networks are quite flexible to design. Another important advantage of queuing network is to model the system in a more detailed level than Markov chain models or fluid flow models. For example, the queuing network in [51] models the P2P system in the peer level, where each peer is modeled as an individual queue. However, the Markov chain model (2) and the fluid flow model (7) (8) are build up at system level, where peers population $x$ and *y* are the interests. Hence queuing network could be extended in many ways [54]. Similar to the accuracy of Markov chain models, the steady state queuing network model is applicable only if a large amount of peers exist in the network. In addition, the assumption of arrival and service process of each queue should be appropriate. The simulation results in [51] validate their queuing network model, where each peer and core router is modeled as an M/G/1/K queue and a GI/G/1 queue, respectively. Queuing network models are not scalable very well due to the limitation of the numerical algorithms solutions. Queuing networks can be divided into two types: (1) Product-form queuing network (PFQN), and (2) Non-product-form queuing network (NPFQN). The PFQN is featured with exponentially distributed inter-arrival and service times. The solution for the steady state probabilities can be expressed as a product of factors describing the state of each node. Although product-form solutions can be expressed easily, considerable computation is needed to analyze even small networks. Hence, instead of exact solution, approximation algorithms are developed to solve large-scale PFQN networks. In order to select a proper approximation algorithm, the trade-off between iteration and accuracy has to be made depending on the particular application [54]. Another more general queuing network is NPFQN, which is feature with non-exponentially distributed service times. Similar to PFQN, approximation algorithms are developed for large-scale NPFQN networks. However, different approximation algorithms are only applicable to solve particular cases of NPFQN networks [54].

### 4.2 Open Issues

Based on the recent modeling work, we notice that the most crucial characteristic of P2P system is dynamics. As we can see in the Table II, only fluid flow models can be applied to model the time-varying performance of the system. However, the shortcoming of fluid flow models is that it might not catch up the details of P2P system due to its typical simple form. As an alternative, queuing network models perform best in capturing the system details and are quite flexible and extendable. Hence, one of the future works could be embedding fluid flow features into queuing network models and evaluating the dynamic performance of P2P system in a detailed model. In addition, the goal of performance modeling is to understand the system and further improve its performance. There are still several weak points in BitTorrent despite of its phenomenal success. The possible improvement could be in the following aspects: (1) modification on the peer selection strategy for seeds to prevent free riding, (2) providing incentive strategy for download bandwidth allocation to improve fairness, (3) ISP-friendly overlay topology design to reduce the cross-ISP traffic.

## 5 Conclusion

In this paper, a broad survey is conducted in the area of performance modeling of BitTorrent P2P file sharing systems. We summarize the majority of modeling techniques into four catalogs: (1) deterministic models, (2) Markov chain models, (3) fluid flow models, and (4) queuing network models. We look into the model with various factors considered for the real BitTorrent system, such as download/upload bandwidth constraints, various link capacity, peer heterogeneous behavior, multiclass peers, bandwidth allocation strategy, impact of free rider, and etc. The performance metrics we focus on include service capacity, download throughput per peer, file download latency, file availability, torrent lifetime and network stability. Furthermore, different modeling techniques are carefully compared on the aspect of applied regime, complexity, accuracy, extensibility, and scalability.

Table I Summary of modeling techniques with factor and metric considered

| Model (typical) | Factors (extension) | Performance Metrics |
|---|---|---|
| Deterministic Model [24] | Multi-piece download [23] [25] | File download latency [23][24][26][27][25] |
| Markov Chain Model [23] $$q((x,y),(x+1,y)) = \lambda$$ $$q((x,y),(x,y-1)) = \gamma y$$ $$q((x,y),(x-1,y+1)) = \mu(\eta x + y)$$ | Download/upload bandwidth constraints [28][29] | Service capacity [23] |
| | | Download throughput per peer [23] |
| | Multiclass peers [30] | File download latency [23][31] |
| | Multi-pieces download [31] | File availability [30][28] |
| | Impact of free rider [32] | Torrent lifetime [29] |
| | Generalized model [20] | Network stability [30][20] |
| Fluid Flow Model [21] $$\frac{dx}{dt} = \lambda - \theta x(t) - \min\{cx(t), \mu(\eta x(t) + y(t))\}$$ $$\frac{dy}{dt} = \min\{cx(t), \mu(\eta x(t) + y(t))\} - \gamma y(t)$$ | Heterogeneous link capacity [36] | Service capacity [40] |
| | Heterogeneous peer behavior [37] | Download throughput per peer [36][38][40][34][21][39] [42] |
| | Bandwidth allocation strategy [38] | |
| | Exponential decreasing peer arrival rate [34] [35] | File download latency [36][38][40][34][21][39][42] |
| | Multiclass peers [39] [40] | File availability [35] |
| | Impact of free rider [41][42][43] | Torrent lifetime [35][28][29] |
| | | Network stability [21][22] |
| Queuing Network Model File Transfer Latency Modeling [51] Peers Behavior Modeling [45] | Heterogeneous peer behavior [45][51] | Service capacity [45] |
| | | Download throughput per peer [45] |
| | Bandwidth allocation strategy [51] | |
| | Multiclass peers [45][51] | File download latency [45][51] |
| | Impact of free rider [45] | |

Table II Comparison of different modeling techniques

|  | Applied Regime | Complexity | Accuracy | Extensibility | Scalability |
|---|---|---|---|---|---|
| Deterministic Model | Initial Transient | Low | Poor | Limited | Very Good |
| Markov Chain Model | Steady state | Low | Conditional | Possible | Limited |
| Fluid Flow Model | Nonstationary | Medium | Good | Good | Good |
| Queuing Network Model | Steady state | High | Conditional | Very Good | Limited |

# References

[1] "Bittorrent," http://www.bittorrent.com/.
[2] "Gnutella," http://www.gnutelliums.com/.
[3] "Freenet," http://www.freenetproject.org/.
[4] I. Stoica, R. Morris, D. Karger, M. F. Kaashoek, and H. Balakrishnan. "Chord: a scalable peerto-peer lookup service for Internet applications," In *Proc. ACM SIGCOMM*, Aug. 2001, pp. 149-160.
[5] S. Ratnasamy, P. Francis, M. Handley, R. Karp, and S. Shenker, "A scalable content-addressable network," in *Proc. ACM SIGCOMM*, Aug. 2001, pp. 161-172.
[6] Hari Balakrishnan, M. Frans Kaashoek, David Karger, Robert Morris, and Ion Stoica, "Looking up data in P2P systems," *Communications of the ACM*, vol. 46, no. 2, pp. 43-48, February 2003.
[7] FastTrack, http://developer.berlios.de/projects/gift-fasttrack/.
[8] Ben Y. Zhao, Yitao Duan, Ling Huang, Anthony D. Joseph and John D. Kubiatowicz, "Brocade: landmark routing on overlay networks," in *Proc. First International Workshop on Peer-to-Peer Systems* (IPTPS), March 2002, pp. 34-44.
[9] "Napster protocol specification," March 2001, http://opennap.sourceforge.net/napster.txt.
[10] M. Izal, G. Urvoy-keller, E. W. Biersack, P. A. Felber, A. A. Hamra, and L. Garces-Erice, "Dissecting BitTorrent: Five months in a torrents lifetime," in *Proc. Passive and Active Measurements* (*PAM*), April 2004, pp. 1-11.
[11] A. Legout, G. Urvoy-Keller, and P. Michiardi, "Rarest first and choke algorithms are enough," in *Proc. 6th ACM SIGCOMM conference on Internet Measurement* (*IMC*), 2006, pp. 203-216.
[12] A. Legout, N. Liogkas, E. Kohler, and L. Zhang, "Clustering and sharing incentives in BitTorrent systems," in *Proc. ACM SIGMETRICS conference on Measurement and Modeling of Computer Systems*, 2007, pp. 301-312.
[13] L. Ellis, "BitTorrent's swarms have a deadly bite on broadband nets," *Multichannel news*, 2006.
[14] T. Karagiannis, P. Rodriguez, and K. Papagiannaki, "Should internet service providers fear peer-assisted content distribution," in *Proc. 5th ACM SIGCOMM conference on Internet Measurement* (*IMC*), 2005, pp. 6-6.
[15] R. Bindal, P. Cao, W. Chan, J. Medved, G. Suwala, T. Bates, and A. Zhang, "Improving traffic locality in BitTorrent via biased neighbor selection," in *Proc. 26th IEEE International Conference on Distributed Computing Systems*, 2006, pp. 66.
[16] C. Gkantsidis and P. Rodriguez, "Network coding for large scale content distribution," in *Proc. IEEE INFOCOM*, 2005, pp. 2235-2245.
[17] D. Levin, R. Sherwood, and B. Bhattacharjee, "Fair file swarming with FOX," in *Proc. 5th International Workshop on Peer-to-peer Systems* (*IPTPS*), 2006.
[18] R. Sherwood, R. Braud, and B. Bhattacharjee, "Slurpie: a cooperative bulk data transfer protocol," in *Proc. IEEE INFOCOM,* vol. 2, March 2004, pp. 941-951.
[19] B. Cohen, "Incentives build robustness in BitTorrent," in *Proc. First Workshop on Economics of Peer-to-peer Systems*, Berkeley, CA, USA, 2003.
[20] Taoyu Li, Minghua Chen, Dah-Ming Chiu, and Maoke Chen, "Queuing models for peer-to-peer systems," in *Proc. 8th International Conference on Peer-to-peer systems* (*IPTPS*), 2009, pp. 4-4.
[21] D. Qiu and R. Srikant, "Modeling and performance analysis of BitTorrent peer-to-peer networks," in *Proc. ACM SIGCOMM*, New York, NY, USA, 2004, pp. 367-378.
[22] Dongyu Qiu and Weiqian Sang, "Global stability of peer-to-peer file sharing systems," *Computer Communications* vol. 31, no. 2, pp. 212-219, Feb. 2008.
[23] X. Yang and G. de Veciana, "Service Capacity of Peer to Peer Networks," in *Proc. IEEE INFOCOM*, vol.4, March 2004, pp. 2242-2252.
[24] G. de Veciana and X. Yang. "Fairness, incentives and performance in peer-to-peer networks," in *Proc. Forty-first Annual Allerton Conference on Communication, Control and Computing*, USA, Oct. 2003.
[25] Xiangying Yang, Gustavo de Veciana. "Performance of peer-to-peer networks: Service capacity and role of resource sharing policies," *Performance Evaluation in P2P Computing Systems*, vol. 63, no. 3,

pp. 175-194, March 2006.
[26] J. Mundinger and R. Weber, "Efficient file dissemination using peer-to-peer technology," Univ. of Cambridge, Cambridge, U.K., Tech. Rep. 2004-01, 2004.
[27] Jochen Mundinger, Richard Weber, and Gideon Weiss. "Analysis of peer-to-peer file dissemination," *SIGMETRICS Performance Evaluation Review*, vol. 34, no. 3, pp. 12-14, Dec. 2006.
[28] R. Susitaival, S. Aalto and J. Virtamo, "Analyzing the dynamics and resource usage of P2P file sharing systems by a spatio-temporal model," in *Proc. International Workshop on P2P for High Performance Computational Sciences* (*P2P-HPCS'06*) *in conjunction with ICCS*, 2006.
[29] R. Susitaival and S. Aalto, "Modeling the population dynamics and the file availability in a BitTorrent p2p system with decreasing Peer Arrival rate," in *Proc. International Workshop on Self-Organizing Systems* (*IWSOS*), 2006.
[30] Y. Tian, D. Wu, and K. W. Ng, "Modeling, analysis and improvement for BitTorrent file sharing networks," in *Proc. IEEE INFOCOM*, April 2006, pp. 1-11.
[31] R. Susitaival and S. Aalto, "Analyzing the file availability and download time in a P2P file sharing system," in *Proc. International Conference on Next Generation Internet Networks*, 2007, pp. 88-95.
[32] Walid Saddi and Fabrice Guillemin, "Measurement based modeling of peer-to-peer file sharing system," in *Proc. 20th International Teletraffic Congress* (*ITC*), 2007, pp. 974-985.
[33] D.G. Luenberger, *Introduction to Dynamic Systems: Theory, Models, and Applications*, New York, NY: Wiley, 1979.
[34] L. Guo, S. Chen, Z. Xiao, E. Tan, X. Ding, X. Zhang, "Measurement, analysis, and modeling of BitTorrent-lik systems," in *Proc. Fifth ACM SIGCOMM conference on Internet Measurement* (*IMC*), 2005, pp. 4-4.
[35] Lei Guo, Songqing Chen, Zhen Xiao, Enhua Tan, Xiaoning Ding, Xiaodong Zhang, "A performance study of BitTorrent-like peer-to-peer systems," *IEEE Journal on Selected Areas in Communications*, vol.25, no.1, pp.155-169, Jan. 2007.
[36] F. Lo Piccolo and G. Neglia, "The effect of heterogeneous link capacities in BitTorrent file sharing systems," in *Proc. International Workshop on Hot Topics in Peer-to-Peer Systems*, 2004, pp. 40-47.
[37] Rivero-Angeles, M.E and Rubino, G., "Priority-based scheme for file distribution in peer-to-peer networks," in *Proc. IEEE International Conference on Communications* (*ICC*), 2010, pp.1-6.
[38] F. Cl´evenot-Perronnin, P. Nain, and K. W. Ross, "Multiclass p2p networks: static resource allocation for service differentiation and bandwidth diversity," *Performance Evaluation*, vol. 62, no. 1-4, pp. 32-49, 2005.
[39] B. Fan, D.-M. Chiu, and J. Lui, "Stochastic analysis and file availability enhancement for BT-like file sharing systems," in *Proc. 14th IEEE International Workshop on Quality of Service* (*IWQoS*), June 2006, pp. 30-39.
[40] B. Fan, Dah-Ming Chiu, and J. Lui, "Stochastic Differential Equation Approach to Model BitTorrent P2P Systems," in *Proc. IEEE International Conference on Communication*, vol.2, 2006, pp. 915-920.
[41] J. Yu, M. Li, F. Hong, and G. Xue, "Free-riding analysis of BitTorrent peer-to-peer networks," in *Proc. IEEE Asia-Pacific Conference on Services Computing* (*APSCC*)*,* Dec. 2006, pp. 534-538.
[42] M. Li, J. Yu and J. Wu, "Free-Riding on BitTorrent Peer-to-Peer File Sharing Systems: Modeling Analysis and Improvement," *IEEE Transactions on Parallel and Distributed Systems*, vol. 19, no. 7, pp. 954-966, July 2008.
[43] J. Yu, M. Li, and J. Wu, "Modeling analysis and improvement for free-riding on BitTorrent file sharing systems," in *Proc. IEEE International Conference on Parallel Processing Workshops* (*ICPPW*)*,* Sept. 2007, pp. 53.
[44] A.G. Werschulz, "Computational complexity of one-step methods for systems of differential equations", Math. Comput., vol. 34, no. 149, pp. 155-174, 1980.
[45] Ge, Z., Figueiredo, D.R., Sharad Jaiswal, Kurose, J., Towsley, D., "Modeling peer-peer file sharing systems," in *Proc. IEEE INFOCOM*, 2003, pp. 2188-2198.
[46] Z. Yao, D. Leonard, X. Wang, and D. Loguinov, "Modeling heterogeneous user churn and local resilience of unstructured P2P networks," in *Proc. 14th IEEE International Conference on Network*

*Protocols* (*ICNP*), Nov. 2006, pp. 32-41.
[47] H. Wu, K. Xu, M. Zhou, A. K. Wong, J. Li, Z. Li, "Multiple-Tree Topology Construction Scheme for P2P Live Streaming Systems Under Flash Crowds", in *Proc. of 2013 IEEE Wireless Communications and Networking Conference (WCNC)*, Shanghai, China, April 2013, pp. 4623-4628.
[48] K. Xu, S. Tipmongkonsilp, D. Tipper, Y. Qian, P. Krishnamurthy "A Time Dependent Performance Model for Multi-hop Wireless Networks with CBR Traffic", in *Proc. of 29th IEEE International Performance Computing and Communications Conference (IPCCC)*, Albuquerque, NM, USA. December 2010, pp. 271 – 280.
[49] K. Xu, D. Tipper, P. Krishnamurthy, and Y. Qian, "An Efficient Hybrid Model and Dynamic Performance Analysis for Multihop Wireless Networks", in *Proc. of 2013 International Conference on Computing, Networking and Communications (ICNC),* San Diego, CA, USA, January, 2013, pp.1090-1096.
[50] K. Xu, D. Tipper, P. Krishnamurthy and Y. Qian, "An Framework of Efficient Hybrid Model and Optimal Control for Multihop Wireless Networks", in *Proc. of 12th ACM International Conference on Measurement and Modeling of Computer Systems (SIGMETRICS'12)*, London, UK, June 2012.
[51] Ramachandran K.K., Sikdar B, "A queuing model for evaluating the transfer latency of P2P systems," *IEEE Transactions on Parallel and Distributed Systems*, vol. 21, no. 3, pp. 367-378, March 2010.
[52] Xia, R.L., Muppala, J.K., "A survey of BitTorrent performance," *IEEE Communications Surveys & Tutorials*, vol.12, no.2, pp.140-158, 2010.
[53] T. D. Dang, R. Pereczes, and S. Moln´ar, "Modeling the population of file-sharing peer-to-peer networks with branching processes," in *Proc. IEEE Symposium on Computers and Communications* (*ISCC*), July 2007, pp.809-815.
[54] Gunter Bolch, Stefan Greiner, Hermann de Meer, Kishor S. Trivedi, *Queueing networks and Markov chains: modeling and performance evaluation with computer science applications*. Wiley-Interscience, New York, NY, 1998.